\documentstyle[prl,aps,floats,epsfig,twocolumn]{revtex}

\begin{document}
\draft

\title{ Muon anomalous magnetic moment in technicolor models} 
         
\author{Zhaohua Xiong $^{a,b,c}$ and Jin Min Yang $^b$} 

\address{$^a$ CCAST (World  Laboratory), P.O.Box 8730, Beijing 100080, China}
\address{$^b$ Institute of Theoretical Physics, Academia Sinica, 
           Beijing 100080, China} 
\address{$^c$ Institute of High Energy Physics, Academia Sinica,
         Beijing 100039, China}
\date{\today}
\maketitle

\begin{abstract}
Contributions to the muon anomalous magnetic moment are evaluated in 
the technicolor model with scalars and topcolor assisted technicolor 
model. In the technicolor model with scalars, the additional contributions
come from the loops of scalars, which were found sizable only for a very 
large $f/f^{'}$ disfavored by the experiment of $b\to s\gamma$.
The topcolor effect is also found to be large only for an unnaturally 
large $\tan\theta'$, and thus the previously evaluated loop effects of extended 
technicolor bosons, suppressed by $m_{\mu}^2/M_{ETC}^2$, must be resorted to account 
for the E821 experiment.  So, if the E821 experiment result persists, it 
would be a challenge to technicolor models.  
\end{abstract} 

\pacs{%PACS:
12.26.NZ, 14.60.Ef, 13.40.Em \quad hep-ph/0102259}
{\bf Key words:} technicolor,  muon,  anomalous magnetic moment 

\bigskip
\noindent{\bf Introduction} 

While it is often argued that the standard model (SM) should be augmented 
by new physics at higher energy scales because of some unanswered fundamental 
questions, the recently reported 2.6 standard deviation of the muon anomalous 
magnetic moment over its SM prediction \cite{E821} may serve as the first 
evidence of existence of new physics at a scale not far above the weak scale
\cite{mar}.  Since the experiment of the muon anomalous magnetic moment will 
be further  developed, it will be a very powerful tool for testing the SM and 
probing  new physics. 

There are numerous speculations on the possible forms of new physics, among 
which supersymmetry (SUSY) and technicolor are the two typical different
frameworks. Although both frameworks are well motivated, SUSY has been more 
favored by precision electroweak experiment than technicolor. The SUSY 
contributions to the muon anomalous magnetic moment were computed by many 
authors\cite{susy1,susy2}. It was found that SUSY can give the large 
contributions and thus naturally explain the reported deviation \cite{susy2}
\footnote{There are also some attempts to explain the reported deviation 
in other approaches\cite{other}.}.

Confronted with the new experiment results of the muon anomalous magnetic moment, 
both supersymmetry and technicolor\cite{tc,tcreview} should be examined. In this 
letter, we will evaluate technicolor contributions. First we will present a 
detailed analysis in the framework of technicolor model with scalars
\cite{Simmons89,Carone94}, which is likely to give significant contributions 
since the couplings of scalars to muon could be significantly enhanced by the 
parameter $f/f^{'}$. Then we give an analysis for the topcolor-assisted technicolor 
model\cite{topcref,tctwohill,tctwoklee}, which also seemingly can give large 
contributions since this model predicts a new gauge boson $Z'$. Finally, for other 
technicolor models, we will give a comment.

\noindent{\bf Technicolor with scalars }

Technicolor with scalars\cite{Simmons89,Carone94}  has a minimal 
$SU(N)$ technicolor sector, consisting  of two techniflavors {\em p} and {\em m}. 
The technifermions transform as singlet under color and as fundamentals under the 
$SU(N)$ technicolor group.  In addition to the above particle spectrum, there 
exists a scalar doublet  $\phi$ to which both the ordinary fermions and technifermions 
are coupled. Unlike the SM Higgs doublet, $\phi$ does not cause electroweak symmetry 
breaking but obtains a non-zero effective vacuum expectation value (VEV)
when technicolor breaks the symmetry.

If we write the matrix form of the scalar doublet as
\begin{equation}
\Phi=\left[ \begin{array}{cc}
            \bar{\phi}^0 & \phi^+\\
            -\phi^-      & \phi^0
            \end{array} \right ]
\equiv \frac{(\sigma+f^{'})}{\sqrt{2}}\Sigma^{'},
\end{equation}
and adopt the conventional non-linear representation $\Sigma=exp(\frac{2i\Pi}{f})$
and $\Sigma^{'}=exp(\frac{2i\Pi^{'}}{f^{'}})$ for technipions, with fields in $\Pi$
and $\Pi^{'}$ representing the pseudoscalar bound states of the technifermions $p$
and $m$,  then the kinetic terms for the scalar fields are  given by
\begin{eqnarray}
{\cal L}_{K.E.}&=&\frac{1}{2}\partial_\mu\sigma\partial^\mu\sigma+
\frac{1}{4}f^2Tr({D}_\mu\Sigma^\dagger {D}^\mu\Sigma)\nonumber\\
&&+\frac{1}{4}(\sigma+f^{'})^2Tr({D}_\mu\Sigma^{'\dagger}{D}^\mu\Sigma^{'}).
\label{kinetic}
\end{eqnarray}
Here $D^\mu$ denotes the $SU(2)_L\times SU(2)_R$ covariant derivative, 
$\sigma$ is an isosinglet scalar field, $f$ and $f{'}$ are the technipion decay 
constant and the effective VEV, respectively.

The mixing between $\Pi$ and $\Pi^{'}$ gives 
\begin{eqnarray} 
\label{pia}
\pi_a&=&\frac{f\Pi+f^{'}\Pi^{'}}{\sqrt{f^2+f^{'2}}}, \\
\label{pip}
\pi_p&=&\frac{-f^{'}\Pi+f\Pi^{'}}{\sqrt{f^2+f^{'2}}},
\end{eqnarray}
with $\pi_a$ becoming the longitudinal component of the W and Z, 
and  $\pi_p$ remaining in the low-energy theory as an isotriplet of physical 
scalars. From Eq.\ (\ref{kinetic}) one can obtain the correct gauge boson masses
providing that $f^2+f^{'2}=v^2$ with the electroweak scale $v=246\ GeV$.

Additionally, the contributions to scalar potential generated by 
the technicolor interactions should be included in this model. The
simplest term one can construct is
\begin{equation}
{\cal L}_T=c_14\pi f^3Tr\left[\Phi\left(
\begin{array}{cc}
h_+ & 0\\
0 & h_-
\end{array}
\right)
\Sigma^\dagger\right] +h.c.,
\label{poential}
\end{equation}
where $c_1$ is a coefficient of order unity, $h_+$ and $h_-$ are the 
Yukawa couplings of scalars to $p$ and $m$. 
From Eq.\ (\ref{poential}) the mass of the charged scalar at lowest order 
is obtained as  
\begin{equation}
m_{\pi_p}^2=2\sqrt{2}(4\pi f/f'){v}^2h
\end{equation}
with $h=(h_++h_-)/2$.  When the largest 
Coleman-Weinberg corrections for the $\sigma$ field are included in 
the effective chiral Lagrangian\cite{Carone94}, one obtains the constraint
\begin{equation}
{\tilde M}_\phi^2 f'+\frac{\tilde \lambda}{2}f^{'3}=8\sqrt{2}\pi c_1hf^3
\end{equation}
and the isoscalar mass as
\begin{equation} 
m_\sigma^2={\tilde M}_\phi^2+\frac{2}{3\pi^2}[6(\frac{m_t}{f'})^4
+Nh^4]f^{'2}
\label{sigma1}
\end{equation}
in limit $(i)$ where the shifted $\phi^4$ coupling $\tilde{\lambda}$ is 
small and can be neglected \cite{Simmons89}, and 
\begin{equation} 
m_\sigma^2=\frac{3}{2}{\tilde \lambda}f^{'2}-\frac{1}{4\pi^2}
[6(\frac{m_t}{f'})^4+Nh^4]f^{'2}
\label{sigma2}
\end{equation}
in limit $(ii)$ where the shifted mass of the scalar doublet $\phi$, 
$\tilde{M_\phi}$ is small and can be neglected\cite{Carone94}.  

The advantage of this model is that it can successfully account for fermion masses 
without generating large flavor-changing neutral current effects, and without 
exceeding the experimental bounds on oblique electroweak radiative corrections. 
Furthermore, we stress at the lowest order, only two independent parameters 
$(f/f^{'},\ m_{\pi_p})$ in the limits (i) and (ii) mentioned above are 
needed to describe the phenomenology.

\begin{center}
\begin{figure}[htb]
\epsfig{file=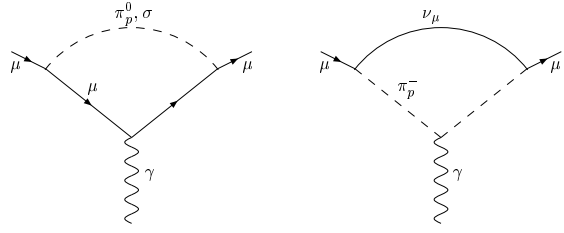 ,width=8cm}              
\caption{ Feynman diagrams for the contributions of  scalars 
           in technicolor with scalars.} 
\label{fey}
\end{figure}
\end{center}

The contributions to the muon anomalous magnetic moment stem from the diagrams 
shown in Fig. \ref{fey}. The relevant interactions can be extracted from 
 Eq.\ (\ref{kinetic}) and  Eq.\ (\ref{poential})\cite{Xiong01}
\begin{eqnarray}
{\cal L}&=&(\frac{v}{f^{'}})\frac{gm_\mu}{2m_W}\sigma\bar{\mu}\mu
-(\frac{f}{f^{'}})\frac{igm_\mu}{2m_W}\pi_p^0\bar{\mu}\gamma_5\mu
\nonumber\\
&&+(\frac{f}{f^{'}})\frac{igm_\mu}
{2\sqrt{2}m_W}\left[\pi_p^+\bar{\nu_\mu}(1+\gamma_5)\mu
-\pi_p^-\bar{\mu}(1-\gamma_5)\nu_\mu\right]\nonumber\\
&&-ieA_\mu\pi_p^{+}\stackrel{\leftrightarrow}{\partial^\mu}\pi_p^- \ .
\end{eqnarray}

Describing the $\gamma\mu\mu$ vertex in the most general Lorentz 
structure form\cite{Jose98} 
\begin{eqnarray}
\Gamma_\alpha&=&i\left\{\gamma_\alpha\left[F_V-F_A\gamma_5\right]
+\left[iF_S+F_P\gamma_5\right]p_\alpha\right.\nonumber\\
&&\left.+\left[iF_M+F_E\gamma_5\right]\sigma_{\alpha\beta}p^\beta\right\}
\label{gamgam}
\end{eqnarray}
with $p$ being the incoming momentum of the photon and the form factors 
$F_i$ being functions of the invariant $s=p^2$, we obtain the
explicit expression of the usual anomalous magnetic dipole moment as 
\begin{eqnarray}
a_\mu &&\equiv\frac{2m_f}{e}F_M(0)=a_\mu^{\pi^0}+a_\mu^{\sigma}
      +a_\mu^{\pi^\pm}\nonumber\\
     &&=\frac{G_Fm_\mu^2}{\sqrt{2}\pi^2}(\frac{f}{f^{'}})^2
     \left\{f_1(a_{\pi_p},b_{\pi_p})+i\frac{m_\mu^2}{m_{\pi_p}}
      f_1(a_{\pi_p},b_{\pi_p})\right.\nonumber\\
     &&\ \ \ \ \ \ \ \ \ \  \  \ \left.+(\frac{v}{f})^2\left[f_1(a_\sigma,b_\sigma)
     +2f_2(a_\sigma,b_\sigma)\right]\right\}.
\label{mumomtc}
\end{eqnarray}
Here $a_h,~b_h$ ($h=\sigma,~\pi_p$) are the
roots of equation $m_\mu^2x^2-m_h^2x+m_h^2=0$ with the convention $a_h>b_h$, 
and the functions $f_i$ are given by
\begin{eqnarray} 
f_1(x,y)&=&\frac{1}{x-y}\left[y^3\ln\frac{y}{y-1}-x^3\ln\frac{x}{x-1}\right]
    +x+y+\frac{1}{2},\nonumber\\
f_2(x,y)&=&\frac{1}{x-y}\left[x^2\ln\frac{x}{x-1}-y^2\ln\frac{y}{y-1}\right]-1.
\end{eqnarray}

\begin{figure}[htb]
\epsfig{file=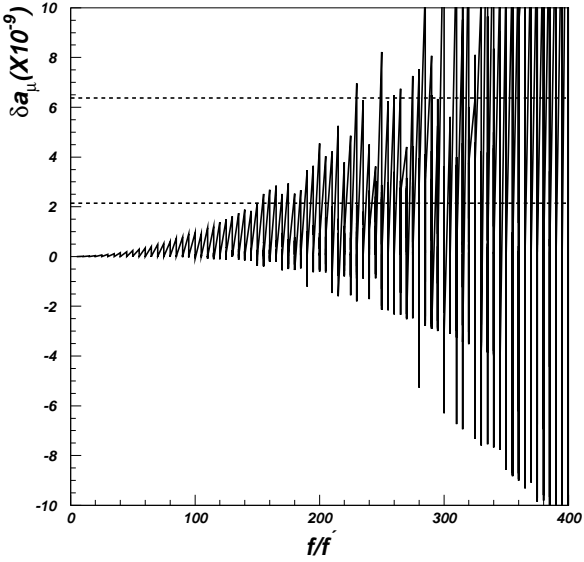,width=8cm}              
\caption{The contributions of the neutral scalars to the muon anomalous 
magnetic moment  as a function of $f/f^{'}$ in the technicolor model with 
scalars.  For  any fixed value of $f/f'$, the mass of $m_{\pi_p}$ is varied 
from 107.7~GeV to 1~TeV. The dashed lines denote the current lower and
upper experimental bounds on new physics contributions at 90\% C.~L.. }
\label{momentr1}
\end{figure}

We should bear in mind that  the parameter space chosen in our numerical calculations
is consistent with known experimental measurements.   At first, we  apply the lower 
experimental bound of $107.7~GeV$ for the SM Higgs and $78.6~ GeV$ for charged Higgs 
boson in Two-Higgs-Doublet model \cite{PDG00} directly to  constrain the mass of the 
scalars, and set the mass of the charged scalar less than $1~TeV$, leaving $f/f^{'}$ 
as a free parameter.  We perform a complete scan of the parameter space of technicolor 
with scalars, and display the additional contribution due to the neutral scalars as 
a function of $f/f^{'}$  in Fig.~\ref{momentr1}, with the mass of 
$m_{\pi_p}$ varying from 107.7~GeV to 1~TeV for any fixed value of $f/f'$.
%For a fixed value of $f/f^{'}$ in Fig.~\ref{momentr1}, 
%the total contribution decreases rapidly as the mass of the scalar increases. 

A couple of remarks are due regarding our results:
\begin{itemize}
\item[{\rm 1.}]
         In limit $(i)$,  contributions to the muon anomalous magnetic moment
         from  the neutral scalars can be positive or negative 
         and not large enough except for very large $f/f^{'}$; 
         whereas the charged scalar can only provide the contribution to the 
         imaginary part which is much smaller than the real one.
\item[{\rm 2.}]         
         Unlike the case in limit $(i)$, the additional contributions are 
         negligible in limit $(ii)$. No very large $f/f^{'}$ is allowed by the 
         model.
\item[{\rm 3.}] 
          As expected,  for any fixed value of $f/f^{'}$, the total contribution 
          decreases rapidly as the mass of the charged scalar increases. 
\end{itemize}
The limit on $f/f^{'}$ has been investigated by several authors, and  obtained from the 
studies of $b\to X_c\tau\nu$ \cite{Xiong99}, $b\to X_s\gamma$\cite{Xiong99,Carone95}, 
$Z\to b\bar{b}$ \cite{Carone95}, $B\to X_s\mu^+\mu^-$ \cite{Su97} and B-$\overline{B}$ 
mixing\cite{Simmons89,Carone94,Carone95}, which is given by
\cite{Xiong99}
\begin{equation}
\frac{f}{f^{'}}\leq 0.03 \left (\frac{m_{\pi_p}}{1\ GeV} \right ) 
                                     \ \ \  (95\%\ C.\ L.).
\label{ffpmpi}
\end{equation}  
According to this limit, $f/f^{'}>200$, which is needed to give large contributions to  
the muon anomalous moment, seems unlikely.

\noindent{\bf  Topcolor-assisted technicolor model} 

The other competitive candidate, which might provide large additional contributions to 
the  muon anomalous moment, is the topcolor-assisted technicolor model
\cite{topcref,tctwohill,tctwoklee}. The model assume: (i) electroweak interactions are 
broken by technicolor; (ii) the top quark mass is large because it is the combination 
of a dynamical condensate component, generated by a new strong dynamics, together with 
a small fundamental component,generated by an extended technicolor (ETC)
\cite{tcreview,etc}; 
(iii) the new strong dynamics is assumed to be chiral critically strong but spontaneously
broken by technicolor at the scale $\sim 1~TeV$, and it generally couples preferentially 
to the third  generation. This needs a new class of technicolor models incorporating 
``top-color''. The dynamics at $\sim 1~TeV$ scale involves the gauge structure:
\begin{eqnarray}
 SU(3)_1&&\times SU(3)_2\times U(1)_{Y_1}\times U(1)_{Y_2}\nonumber\\
 &&\rightarrow SU(3)_{QCD}\times U(1)_{EM}\nonumber
\end{eqnarray}
where $SU(3)_1\times U(1)_{Y_1}~[SU(3)_2\times U(1)_{Y_2}]$ generally couples  
preferentially to the third (first and second) generation, and is assumed to be strong
enough to form chiral $<\bar{t}t>$ but not $<\bar{b}b>$ condensation by the 
$U(1)_{Y_1}$ coupling.  A residual global symmetry $SU(3)^{'}\times U(1)^{'}$ 
implies the existence of a massive color-singlet heavy $Z^{'}$ and an octet $B_\mu^A$.
A symmetry-breaking pattern onlined above will generically give rise to three
top-pions, $\tilde{\pi}$, near the top mass scale.       

In this model, in addition to the previously evaluated loop effects of 
ETC bosons which generates the muon mass \cite{mar}, the muon anomalous 
magnetic moment  receives additional contributions 
only from a gauge boson $Z_\mu^{'}$, top-pions ${\tilde{\pi}}^0,~{\tilde{\pi}}^\pm$
and technipions with Feynman diagrams similar to  Fig.~\ref{fey} and $\sigma,~\pi_p$ 
replaced by  $Z_\mu^{'}$, top-pions (technipions), respectively.
 The interaction of top-pions, neutral gauge boson with muon are given 
by \cite{tctwohill} 
\begin{eqnarray}
{\cal L}^{eff}&=&\frac{1}{2}g_1\tan^2\theta^{'}Z^{'}_\alpha
\left[\bar{\mu}_L\gamma^\alpha\mu_L+2\bar{\mu}_R\gamma^\alpha\mu_R\right]\nonumber\\
&&+\frac{m_\mu}{f_\pi}\left[\frac{i}{\sqrt{2}}\bar{\mu}_L\tilde{\pi}^0\gamma_5\mu_R+
\bar{\nu}_{\mu L}\tilde{\pi}^+\mu_R+h.c.\right]
\label{topcl}
\end{eqnarray}
where $g_1$ is the $U(1)_Y$ coupling constant at the scalar $\sim 1~TeV$. 
The SM $U(1)_Y$ and the $U(1)^{'}$ field $Z_\alpha^{'}$ is then defined by 
orthogonal rotation with mixing angle $\theta^{'}$.  

In this letter, we  don't take small technipions effects into account. Note    
when the Lagrangian in Eq.~(\ref{topcl}) is used to calculate the additional 
technicolor effects to the muon anomalous magnetic moment, the small contribution from 
the charged top-pions, as well as neutral one, can be neglected safely, as in the case 
of technicolor with scalars. Now we present the contributions from  the neutral 
gauge boson $Z^{'}$ 
\begin{eqnarray}
a_\mu &=&\frac{g_1^2}{32\pi^2}\tan^2\theta^{'}
      \left\{-a_{Z^{'}}-b_{Z^{'}}-\frac{11}{2}\right.\nonumber\\
      &&\left.+\frac{1}{a_{Z^{'}}-b_{Z^{'}}}
       \left[a_{Z^{'}}(a_{Z^{'}}^2+5a_{Z^{'}}-6)\ln\frac{a_{Z^{'}}}{a_{Z^{'}}-1}
      \right.\right.\nonumber\\
      &&\left.\left.-b_{Z^{'}}(b_{Z^{'}}^2+5b_{Z^{'}}-6)
       \ln\frac{b_{Z^{'}}}{b_{Z^{'}}-1}\right]\right\},
\end{eqnarray}
where $a_{Z^{'}},~b_{Z^{'}}$ are defined in Eq.~(\ref{mumomtc}).
\begin{figure}[htb]
\epsfig{file=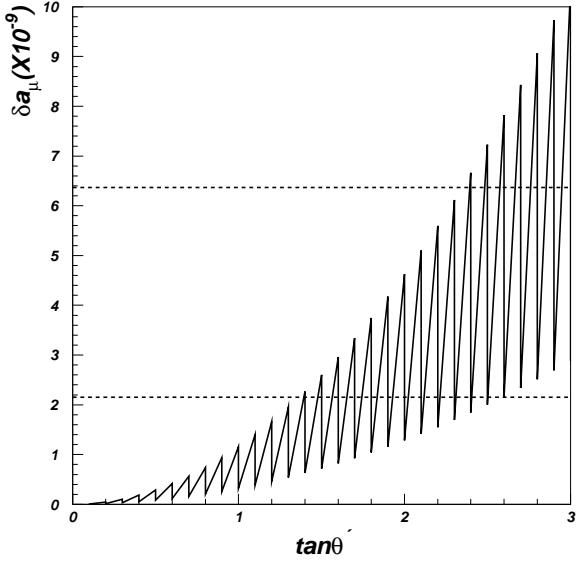,width=8cm}              
\caption{Additional contribution due to the neutral gauge boson $Z^{'}$  
to the muon anomalous magnetic moment  as a function of $\tan\theta^{'}$.
For any fixed value  of $\tan\theta^{'}$, the mass of $Z^{'}$ is varied 
from 100~GeV to 1~TeV. The dashed lines are the same as in Fig.~\ref{momentr1}.}
\label{moment}
\end{figure}

The additional contribution due to the neutral gauge boson $Z^{'}$  
to the muon anomalous magnetic moment  as a function of $\tan\theta^{'}$ is shown
is Fig.~\ref{moment}, with the mass of $Z^{'}$ varying from 100~GeV to 1~TeV
for any fixed value  of $\tan\theta^{'}$. One can see the topcolor-assisted model 
can also give large contributions in case of a large $\tan\theta'$, which, however, 
is not favored by the model because a small $\tan\theta^{'}\ll 1$ is ultimately 
demanded to select the top quark direction  for condensation.
To account for the E821 experiment, the ETC loop contributions must be resorted, 
which is suppressed by $m_{\mu}^2/M_{ETC}^2$ and  requires $M_{ETC} \sim 1$ TeV,  
as evaluated in \cite{mar}.

We also scanned other technicolor models and found that  
models without ETC to generate the fermions masses are more unlikely to 
give the large contributions to the muon anomalous magnetic moment since the 
new particle couplings to the muon do not have any enhancement factors like 
$f/f'$.    
 
In summary, we evaluated technicolor contributions to the muon anomalous magnetic 
moment in two frameworks of technicolor: technicolor model with scalars and 
the topcolor-assisted technicolor model. 
We found that the technicolor model with scalars can give the large 
contributions required by the E821 experiment only for a very large $f/f^{'}$ which, 
however, is disfavored by the experiment of $b\to s\gamma$. 
The $Z'$ loop in the topcolor-assisted  model can also give large contributions in
case of a large $\tan\theta'$,  which is not in accord with the motivation for 
building this model. Thus to account for the E821 experiment, the ETC loop 
contributions suppressed by $m_{\mu}^2/M_{ETC}^2$ \cite{mar} must be resorted. 

Therefore, if the current deviation of the muon anomalous moment from its SM 
prediction persists as the experiment is further developed, it would be a 
challenge to the technicolor models. 

\noindent {\bf Acknowledgments} 

We thank L. Y. Shan, C. X. Yue and X. Zhang for helpful discussions.
This work is supported in part by a grant of 
Chinese Academy of Science for Outstanding Young Scholars.

\end{document}